\documentclass[prb,twocolumn,superscriptaddress,floats]{revtex4}

\usepackage[dvips]{graphicx}
\usepackage{amsmath}
\usepackage{booktabs}
\usepackage{dcolumn}
\usepackage{sidecap}

\usepackage{rotating}

\newcolumntype{d}{D{.}{.}{-1}}

\begin{document}


\title{Strong Magneto-elastic coupling in VOCl: Neutron and synchrotron powder x-ray diffraction study}

\author{A. C. Komarek }
\affiliation{II. Physikalisches Institut, Universit\"{a}t zu
K\"{o}ln, Z\"{u}lpicher Str. 77, D-50937 K\"{o}ln, Germany}
\author{T. Taetz}
\affiliation{Institut f\"{u}r Anorganische Chemie, Universit\"{a}t zu
K\"{o}ln, Greinstr. 6, D-50939 K\"{o}ln, Germany}
\author{M.T. Fern\'andez-D\'iaz}
\affiliation{Institut Laue-Langevin, 38042 Grenoble, France}
\author{D. M. Trots}
\affiliation{Hasylab/DESY, Notkestr. 85, D-22607, Hamburg,
Germany}
\author{A. M\"{o}ller}
\affiliation{Institut f\"{u}r Anorganische Chemie, Universit\"{a}t zu
K\"{o}ln, Greinstr. 6, D-50939 K\"{o}ln, Germany}
\author{M. Braden}
\email{braden@ph2.uni-koeln.de}
\affiliation{II. Physikalisches
Institut, Universit\"{a}t zu K\"{o}ln, Z\"{u}lpicher Str. 77,
D-50937 K\"{o}ln, Germany}

\date{\today}

\pacs{PACS numbers:}


\begin{abstract}

We present a combined neutron and synchrotron powder diffraction
study of the crystal and magnetic structure of VOCl. The
occurrence of antiferromagnetic order in VOCl is accompanied by a
monoclinic distortion. The high sensitivity of the magnetic
interaction parameters on V-O-V bond angles and V-V distances
yields strong magnetoelastic coupling which implies the structural
distortion below T$_N$ as well as anomalous structural effects in
the paramagnetic phase.

\end{abstract}
\maketitle

\section{Introduction}
The transition metal oxyhalides MOX with M = Sc, Ti, V, Cr, Fe
and X = Cl, Br have attracted strong interest in the past years
initiated through the  challenging low-dimensional phenomena
observed for  TiOCl
\cite{tioclshaz,tioclseidel,tioclkataev,tioclrueckamp,tioclcaimi,tioclschoenleber}.
At T$_1$ = 67 K, TiOCl shows a transition associated with a
dimerization of the Ti-ions along the $b$-direction (first order
phase transition) which has been interpreted as a spin Peierls
transition. Thus, TiOCl might be a rare case of an inorganic spin
Peierls system similar to CuGeO$_3$
\cite{tioclshaz,tioclseidel,cugeo1}. The whole group of transition
metal oxyhalides MOX provides an interesting class of -- at
room-temperature --  iso-structural compounds in which several
electronic configurations and their interactions with the crystal
lattice can be studied systematically. However, the physical
properties of the MOCl compounds different from TiOCl have been
studied in less detail so far
\cite{tioclvenien,tioclwiedenmann,feocl,feoclB,crocl}. Like
TiOCl, VOCl crystallizes in an orthorhombic structure with space
group $Pmmn$, with a single crystallographic V-site
\cite{voclhaase}. The V$^{3+}$-ions are coordinated by four
oxygen atoms and two chlorine atoms forming strongly distorted
VO$_4$Cl$_2$ octahedrons; see Fig. \ref{figI}. These are linked
via corners and edges forming quasi-two-dimensional bilayers in the
$ab$-plane. Due to the Cl- ions there is a clear separation of
neighboring Ti-O-bilayers explaining the electronic and magnetic
low-dimensional character. Due to electronic correlations all
oxyhalides are Mott-insulators. In VOCl the band gap is reported
to be about 1.7 eV \cite{tioclbenckiser}. Whereas a direct overlap
of the ground state $yz$-orbitals (with reference to the cell axis
$b$ and $c$) along the
crystallographic $b$-axis leads to one dimensional magnetic
interactions in TiOCl \cite{tioclrueckamp}, VOCl is believed to
exhibit quasi-two-dimensional antiferromagnetic ordering below
T$_N$ (80 K) \cite{tioclwiedenmann}. In an early powder neutron
diffraction study Wiedenman \emph{et al.} reported a collinear
magnetic order with the wave vector (1/2 1/2 1/2) and the magnetic
moments pointing in $a$-direction. The magnetic susceptibility
was well described within a quadratic 2D-Heisenberg model.
However, no signs of a structural transition in VOCl have been
observed at low temperatures \cite{tioclwiedenmann}.

\begin{figure}[!ht]
\begin{center}
\includegraphics*[width=1\columnwidth]{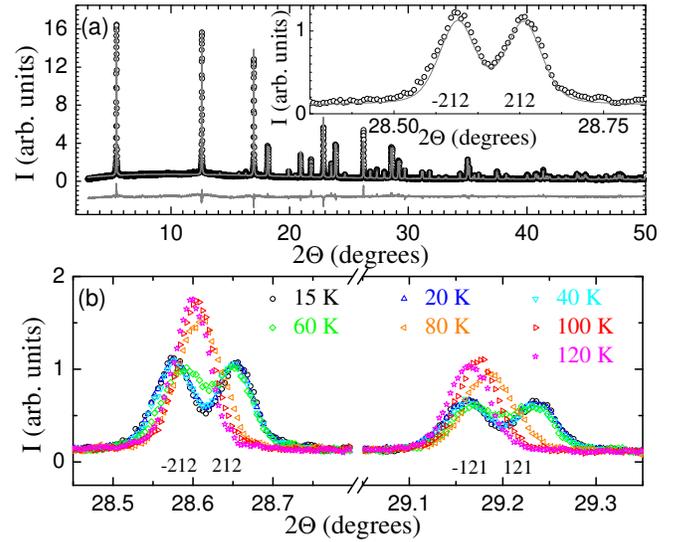}
\end{center}
\caption{(color online) Synchrotron radiation powder x-ray
diffraction patterns. (a) Rietveld fit (\emph{gray}) to the data
at 15 K (\emph{black}) with an enlargement of the diffraction
pattern around the ($\pm$2 1 2) profiles in the inset. (b)
Temperature dependency of the ($\pm$212) and ($\pm$121)
profiles.}\label{figA}
\end{figure}

In this work, we report x-ray and neutron powder diffraction
experiments on VOCl which reveal a structural phase transition
accompanying the antiferromagnetic order in  VOCl. We find a
remarkably strong magnetoelastic coupling in VOCl which seems to
arise from the dependence of the magnetic interaction on soft
lattice degrees of freedom, V-O-V bond angles and V-V distances,
in analogy to the spin-Peierls compound CuGeO$_3$
\cite{cugeo1,cugeo2,cugeo3}.

\section{Experimental}

VOCl was prepared by the chemical transport technique as described
by Sch\"{a}fer et al. \cite{tioclschafer}. Single crystals
obtained via this route were thoroughly ground, compressed under
hydrostatic pressure and reground afterwards. This procedure was
repeated several times in order to achieve a homogenous powder
and hence minimize preferred orientation effects. The phase
purity of the sample was checked several times using powder x-ray
diffraction.

Synchrotron radiation powder x-ray diffraction measurements have
been performed at beamline B2 at Hasylab/DESY in Hamburg,
Germany \cite{beamlineB2A}. The single phase VOCl-powder was finely ground and filled in
a glass capillary with 0.3 mm in diameter.  The incident X-ray
wavelength was 0.7466 \AA\ and the reflections have been measured
using an image plate detector in Debye-Scherrer geometry \cite{beamlineB2B}.

Powder neutron diffraction measurements have been performed at
the D1A, D2B and D20 diffractometer at the ILL in Grenoble,
France ($\lambda$~=~1.91, 1.594 and 2.4233 \AA, respectively). The
samples measured at the D20 and D2B diffractometer contained some
unidentified impurity phases. Therefore, only the single phase
sample measured at the D1A diffractometer was used for a full
nuclear-structure refinement.  The high-intensity D20 data have
been used for the magnetic structure refinement, as there was no
overlap of peaks from the impurity phases with the magnetic peaks
due to the longer wave length.

\begin{figure}[!ht]
\begin{center}
\rotatebox{0}{\includegraphics*[width=1\columnwidth]{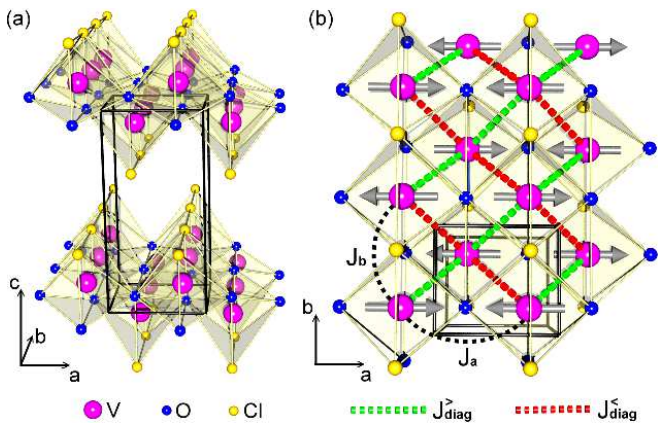}}
\end{center}
\caption{(color online) Nuclear and magnetic structure of VOCl;
\emph{yellow}: Cl-, \emph{blue}: O-, \emph{magenta}: V-ions;
\emph{black}: the structural unit cell. The directions of the
magnetic moments at each V-site is indicated by \emph{magenta
arrows}. \emph{Green} (\emph{red}) \emph{dotted lines} indicate
the antiferromagnetically (ferromagnetically) coupled V-V pairs in
the AFM1 structure.} \label{figI}
\end{figure}

\section{Results and Discussion}

\subsection{Monoclinic distortion}

\begin{figure}[!t]
\begin{center}
\includegraphics*[width=0.95\columnwidth]{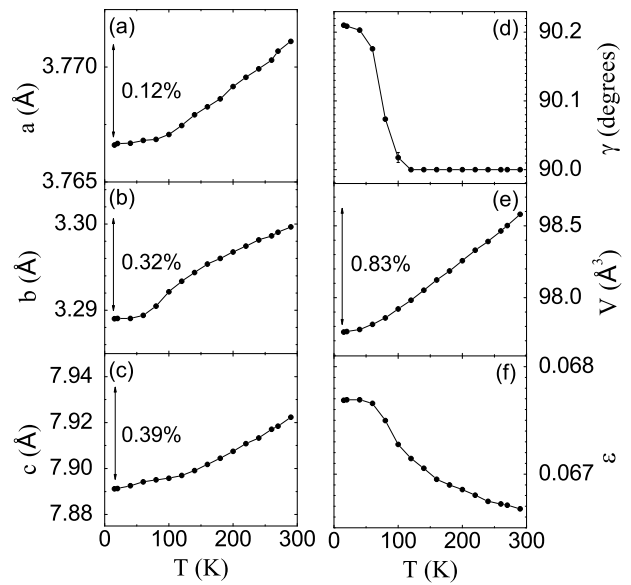}
\end{center}
\caption{Results of synchrotron  radiation powder x-ray
diffraction measurements: (a-c) lattice parameter, (d) monoclinic
angle, (e) unit cell volume and (f) orthorhombic splitting
$\varepsilon$~=~(a-b)/(a+b). (Lines are guide to the eyes.)}
\label{figB}
\end{figure}

The temperature dependent powder x-ray diffraction measurements
using synchrotron radiation clearly reveal a structural phase
transition below 80 K. A diffraction pattern at the lowest
temperature, 15 K, is shown in Fig. \ref{figA} (a). The open
circles denote the measured data points and the line indicates
the Rietveld fit of the structure model using the program
\emph{FullProf/WinPLOTR} \cite{fullprof}. The part of the
diffraction pattern around the ($\pm$212) reflections is shown in
the inset. The clearly observable splitting of the orthorhombic
(212) reflection indicates the symmetry reduction from the
orthorhombic ($Pmmn$) to a monoclinic symmetry. The data can be
well described in space group $P2/n$ (monoclinic setting $c$) and
a monoclinic angle $\gamma$ of 90.2\ensuremath{^\circ}. In Fig.
\ref{figA} (b) the ($\pm$212) and ($\pm$121) reflections are
plotted for different temperatures indicating the orthorhombic to
monoclinic phase transition below 80 K. The resulting lattice
parameters are plotted as a function of temperature in Fig.
\ref{figB} (a-d). A rise of the monoclinic angle $\gamma$ can be
observed below 80 K reaching its maximum value of about
90.2\ensuremath{^\circ} around 40 K. The monoclinic distortion
and the structural phase transition apparently have escaped
detection in the previous neutron diffraction experiments due to
their limited resolution \cite{tioclwiedenmann}. The
group-subgroup relationship between $Pmmn$ and $P2/n$ as well as
the small change in the unit cell volume suggest a second order
phase transition.

\begin{figure}[!ht]
\begin{center}
\includegraphics*[width=1\columnwidth]{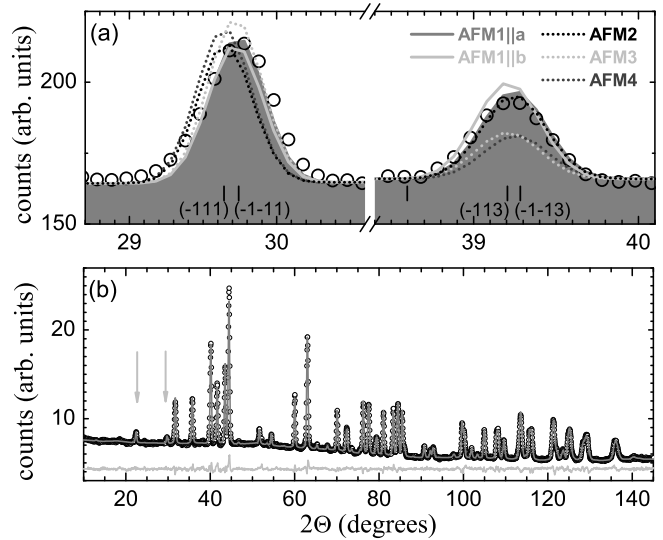}
\end{center}
\caption{(a) Powder neutron diffraction patterns  measured at 2~K
(D20 diffractometer); \emph{circles}: data points, \emph{lines}:
Rietveld fits to the data with the AFM1-AFM4 structure models.
The area below the AFM1-fit (moment $\|a$) is filled
(\emph{gray}). (b) Fit of the D1A measurement at 10~K;
\emph{black}: data points, \emph{gray}: Rietveld fit to the data,
\emph{lt. gray}: difference I$_{obs}$-I$_{calc}$, arrows indicate
the magnetic peaks.} \label{figC}
\end{figure}

Powder neutron diffraction measurements have been performed at
the ILL. In Fig. \ref{figC} (a-b) neutron diffraction patterns of
VOCl are shown.  Fig. \ref{figC} (a) shows the magnetic peaks
measured at 2~K at the high-flux D20 diffractometer with an
incident wavelength of 2.4233~\AA. In Fig. \ref{figC} (b) an
entire diffractogram measured at 10~K on the D1A diffractometer
using a neutron wavelength of 1.91~\AA\ and higher resolution is
shown. The latter data was used for structural refinements.

The appearance of magnetic peaks at (1/2~1/2~1/2) and
(1/2~1/2~3/2) with respect to the nuclear lattice clearly
indicates the antiferromagnetic order occurring below T$_N\sim$80K
in perfect agreement with Ref. \cite{tioclwiedenmann}. A fully
three-dimensional (3D) antiferromagnetic structure could be
fitted to the neutron data; see Fig. \ref{figC} (a). The
half-indexed reflections indicate the magnetic propagation vector
$\mathbf{k}$ $=$ (1/2~1/2~1/2) corresponding to a
$2\times2\times2$ enlargement of the magnetic unit cell.

Within the nuclear unit cell there are two magnetic sites, the V
ions at (0.75,0.25,0.117) and at (0.25,0.75,-0.117) which belong
to two different chains of edge-sharing VO$_4$Cl$_2$ octahedrons
along the $b$-direction, labelled $b$-chains in the following.
These $b$-chains form the VO-planes parallel to the $a,b$-planes
and are situated alternatingly above and below the z=0 level. The
inspection of the magnetic paths in Fig. 2 (a) and (b) allows one
to qualitatively understand the antiferromagnetic order. There is
a magnetic interaction $J_a$ along $a$ mediated through the common
O between two V sites at the same $z$-level.  This
antiferromagnetic interaction, which is absent in TiOCl due to the
single occupation of the $d_{yz}$ orbital, only acts on V-ions in
next-nearest neighbor $b$-chains yielding an antiferromagnetic
arrangement between these next-nearest $b$-chains and thus a
doubling of the $a$ parameter. There is also an interaction $J_b$
along $b$ mediated through the common octahedron edges which
results in an antiferromagnetic order within the $b$-chains and
doubling along $b$. With an additional weak antiferromagnetic
coupling perpendicular to the planes one may thus explain the
doubling of the magnetic lattice along the three orthorhomic
directions. However, in the orthorhombic phase the interaction
between the nearest-neighbor $b$-chains is fully frustrated since
the path parallel to [110] and that parallel to [1-10] are
equivalent by symmetry and thus $J_{diag}^{>}$=$J_{diag}^{<}$.
Since these two interaction parameters connect to spins of
opposite sign in the nearest-neighbor $b$-chain the coupling is
fully frustrated in the orthorhombic phase.

The two possible magnetic arrangements in the orthorhombic phase,
however, do not correspond to two distinct magnetic symmetries but
only to two domains of equivalent magnetic symmetry. This
conclusion is fully corroborated by the representation analysis of
the orthorhombic structure which only yields one irreducible
representation for the propagation vector $\mathbf{k}$ $=$
(1/2~1/2~1/2) and magnetic moments aligned parallel to the
$a,b$-planes. The mono\-clinic distortion accompanying the
magnetic order lifts the frustration of the nearest-neighbor
$b$-chain coupling, $J_{diag}^{>}\ne J_{diag}^{<}$ , since it
renders the magnetic paths parallel to [110] and [1-10]
inequivalent. As indicated in Fig. 2 (b), all the $J_{diag}$-paths
parallel to [110], $J_{diag}^{>}$, are equivalent in the
monoclinic phase but they differ from the $J_{diag}$-paths
parallel to [1-10], $J_{diag}^{<}$. Supposing that $J_{diag}^{>}$
is larger than $J_{diag}^{<}$ the magnetic ordering shown in Fig.
2 (b) is stabilized. We may thus conclude that the monoclinic
distortion accompanying the magnetic transition lifts the
frustration of the magnetic coupling between nearest-neighbor
$b$-chains.

\subsection{Magnetic structure}

Let us now turn to the quantitative analysis of the magnetic
structure in VOCl. The two strong magnetic Bragg peaks observed at
(1/2~1/2~1/2) and (1/2~1/2~3/2) are best described by the model
with ferromagnetic coupling between the two V-moments at
(0.75,0.25,+z) and  (0.25,0.75,-z) pointing in $a$-directipon;
see Fig. \ref{figI}. Other
directions of the magnetic moment like the [110]-direction induce
discrepancies in the measured intensities of the ($\pm$1$\pm$11)
and ($\pm$1$\pm$13) reflections (in notation of the magnetic unit
cell).

Due to representation theory, there are four irreducible
representations for space group $P 1 1 2/n$ with a magnetic ion
at (-0.25~0.25~$z_0$) ($z_0$~$\approx$~0.117) and with propagation
vector $\overrightarrow{k}$~=~(1/2~1/2~1/2). The basis vectors
for these four possibilities derived from the program $BasiRep$
\cite{fullprof} are shown in Fig. 5. These four possibilities are
referred to as the antiferromagnetic structures AFM1 to AFM4. The
representation analysis separates the arrangements with the
moment parallel to $c$ (AFM3 and AFM4) from those with in-plane
moments (AFM1 and AFM2). For these two cases, there is the
possibility of ferromagnetic or antiferromagnetic coupling of the
two sites at (0.75,0.25,+z) and  (0.25,0.75,-z). The two magnetic
structures AFM1 and AFM2 (or AFM3 and AFM4) can not be
distinguished if the crystal structure is orthorhombic ($Pmmn$).
However, the structural phase transition $Pmmn$ $\rightarrow$
$P2/n$ observed in our work induces a splitting of the ($\pm$1
$\mp$1 $1$) and the ($\pm$1 $\pm$1 $3$) reflections (referring to
the magnetic cell). The small splitting of these reflections
allows us to distinguish between the two magnetic structures. In
Fig. \ref{figC}~(a) the Rietveld fits to the measured data at 2~K
are shown for the four magnetic models. In the Rietveld fit of
the AFM2 structure, the calculated intensities for the first
magnetic peak ($\pm$1~$\pm$1~1) are shifted to lower $2\Theta$
values in contradiction to the experiment.  In contrast, the AFM1
structure perfectly describes the data if the magnetic moments
are parallel to the $a$-direction. A fit with moments in $b$
direction does not result in a correct description of the
intensities. Even the free refinement of both $a$- and
$b$-components yields only negligible values for the moment in
$b$-direction. This observation of the ordered moment along the
$a$ direction is in accordance with the measured magnetic
susceptibility in Ref. \cite{tioclwiedenmann} which sharply drops
along the $a$-direction below $T_N$. The AFM3 and AFM4 structures
with $c$-components do not properly describe the magnetic
intensities; see Fig. \ref{figC}~(a). In VOCl the AFM1 magnetic
structure with moments in $a$-direction is thus realized. In Ref.
\cite{tioclwiedenmann}, a mixture of the AFM1 and AFM2 was
proposed basing on single-crystal diffraction, however, most
likely, the twinning induced by the structural phase transition,
which was not detected in that work, occurs in the used single
crystal fully explaining the observed intensities.

The antiferromagnetic moment amounts to about 1.3 $\mu_B$ at 2 K
which is in approximate agreement with the value reported in Ref.
\cite{tioclwiedenmann} but which is  smaller than the full spin
moment expected for the vanadium ion with a 3d$^2$ electronic
configuration. The value of the antiferromagnetic moment as a
function of temperature is shown in Fig.~\ref{figD}~(h). The
temperature dependency of this antiferromagnetic moment indicates
a N\'{e}el-temperature of T$_N\sim$80~K.

\begin{figure}[!ht]
\begin{center}
\includegraphics*[width=1\columnwidth]{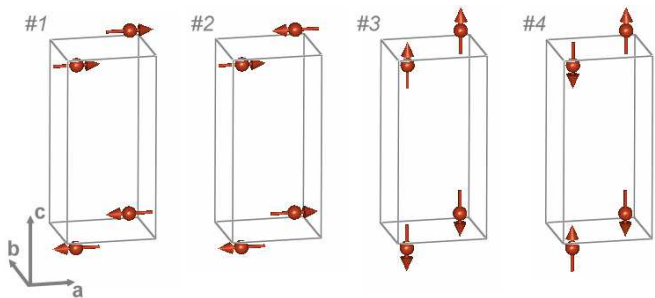}
\end{center}
\caption{(color online) (a-d) The possible irreducible
representations in space group $P 1 1 2/n$ for k~=~(1/2~1/2~1/2)
with magnetic atoms in the primitive unit cell at (-0.25, 0.25,
$z_0$). For (a) and (b) the presented spin direction along $a$
indicates the direction of the ordered moment found in VOCl, which
exhibits magnetic order according to the representation \# 1.}
\label{figIR}
\end{figure}

\subsection{Magnetoelastic coupling in VOCl}

\begin{figure}
\begin{center}
\includegraphics*[width=0.95\columnwidth]{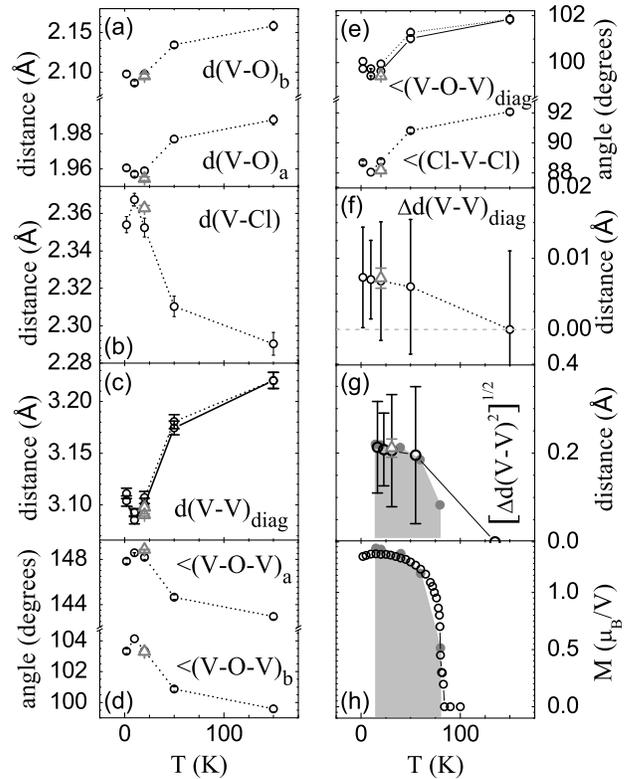}
\end{center}
\caption{Results of powder neutron diffraction  (\emph{black
circles}) and synchrotron measurements (\emph{gray triangles}).
(a) V-O distances for bonds in $a$- and $b$-direction. (b) V-Cl
distance. (c) V-V distances for V-ions within the same V-bilayer
in diagonal direction (\emph{red}  and \emph{green} bonds in Fig.
\ref{figI}). (d) V-O-V bond angles for V-ions (with the same value
of $z$) in $a$-direction (V-O-V)$_a$ and in $b$-direction
(V-O-V)$_b$. (e) Cl-V-Cl angle and V-O-V angle in diagonal
direction. (f) $\Delta$d(V-V)$_{diag}$ denotes the difference of
the two diagonal V-V distances. (g)
$\sqrt{\{d(V-V)_{diag,1}\}^2-\{d(V-V)_{diag,2}\}^2}$ plotted with
the monoclinic angle $\gamma$ (\emph{gray}). (h) The
antiferromagnetic moment (\emph{black}) together with $\gamma$
(\emph{gray}). (Lines are guide to the eyes.)} \label{figD}
\end{figure}

The structural refinements reveal a displacement of the vanadium
ions away from the chlorine towards the oxygen sites in
$c$-direction upon cooling, see Fig. \ref{figD} (a-d). The
VO-planes become thus flatter at low temperature. In consequence
the V-O-V bond angle along $a$ increases upon cooling thereby
enhancing the antiferromagnetic exchange interaction $J_a$. The
V-O-V path along $a$ exhibits a bond angle considerably smaller
than 180\ensuremath{^\circ} but larger than the value of
90\ensuremath{^\circ} where one expects the antiferromagnetic
superexchange to vanish \cite{cugeo1,cugeo2,cugeo3}. In
consequence the magnetic interaction $J_a$ depends sensitively on
the bond angle and increases with its enhancement. The direct
overlap of the V orbitals along $b$ seem to dominate the magnetic
interaction $J_b$ which increases upon cooling due to the
shrinking of the $b$-parameter. The structural changes implied
through the cooling appear thus to be driven by the magnetism
which favors an enhancement of the magnetic interaction
parameters at low temperature.

Magnetoelastic coupling is also evident in the coupling of the
magnetic transition with the monoclinic distortion. In
Fig.~\ref{figD}~(h) we plot the ordered magnetic moment together
with the monoclinic distortion as a function of temperature
demonstrating the close coupling.  The structural phase transition
from $Pmmn$ to $P112/n$ symmetry is of proper ferroelastic
character, so that the monoclinic distortion is the primary
structural order parameter.

\begin{table}[!t]
\centering{ {\footnotesize
\begin{ruledtabular}
\begin{tabular}[t]{llllllll}
\textbf{2 K:} & neutron & D1A (ILL)\\
\\
\emph{SG}                & \textbf{a (\AA)}   & \textbf{b (\AA)}    & \textbf{c (\AA)}      &  $\mathbf{\gamma}$            \\
\textbf{$P 1 1 2/n$}      &   3.76755(10)       & 3.29045(9)         & 7.89582(20)        & 90.21(0)         \\
\hline
\emph{ Atom}  & \textbf{x}& \textbf{y}    & \textbf{z}      &               \\
\hline
\textbf{V}      &   -0.25    &  0.25        &  0.11678(76)   &                \\
\textbf{Cl}     &   -0.25    &  0.75        &  0.33000(4)    &                \\
\textbf{O}      &   -0.25    &  0.75        & -0.04800(6)    &                \\
\hline
 \\ \\
\hline
\textbf{10 K:} & neutron & D1A (ILL)\\
\\
\emph{SG}                & \textbf{a (\AA)}   & \textbf{b (\AA)}    & \textbf{c (\AA)}      &  $\mathbf{\gamma}$            \\
\textbf{$P 1 1 2/n$}      &   3.76776(6)     & 3.29076(5)         & 7.89611(13)        & 90.202(1)        \\
\hline
\emph{ Atom}  & \textbf{x}& \textbf{y}    & \textbf{z}      &               \\
\hline
\textbf{V}      &   -0.25    &  0.25        &  0.11478(59)   &                \\
\textbf{Cl}     &   -0.25    &  0.75        &  0.33035(3)    &                \\
\textbf{O}      &   -0.25    &  0.75        &  -0.04765(3)    &                \\
 \\ \\
\hline
\textbf{150 K:} & neutron & D1A (ILL)\\
\\
\emph{SG}                & \textbf{a (\AA)}   & \textbf{b (\AA)}    & \textbf{c (\AA)}      &  $\mathbf{\gamma}$            \\
\textbf{$P m m n S$}      &   3.77001(10)     & 3.29656(10)         & 7.90713(25)        & 90         \\
\hline
\emph{ Atom}  & \textbf{x}& \textbf{y}    & \textbf{z}      &               \\
\hline
\textbf{V}      &    -0.25    &  0.25          &   0.12804(111)   &                \\
\textbf{Cl}     &    -0.25    &  0.75          &   0.32914(5)    &                \\
\textbf{O}      &   -0.25    &  0.75           &  -0.04815(8)    &                \\   
 \\ \\
\hline
\textbf{20 K:} & synchrotron & B2 (DESY)\\
\\
\emph{SG}                & \textbf{a (\AA)}   & \textbf{b (\AA)}    & \textbf{c (\AA)}      &  $\mathbf{\gamma}$            \\
\textbf{$P 1 1 2/n$}      &   3.7654(3)     & 3.2880(2)         & 7.8894(5)       & 90.207(1)         \\
\hline
\emph{ Atom}  & \textbf{x}& \textbf{y}    & \textbf{z}      &               \\
\hline
\textbf{V}      &   -0.25    &  0.25        &   0.11555(15)   &                \\
\textbf{Cl}     &   -0.25    &  0.75        &   0.33068(19)    &                \\
\textbf{O}      &   -0.25    &  0.75        &  -0.04906(51)    &                \\
\end{tabular}
\end{ruledtabular}
}} \caption{\label{tab1} Results of the structural refinement of
the neutron (D1A) and synchrotron (B2) measurements. }
\end{table}

More quantitatively, the preference of the AFM1 structure can be
understood with regard to the monoclinic distortion of the
structure. As noted above, the magnetic arrangement is
antiferromagnetic along the V-V bonds, $J_{diag}^{>}$
(\emph{green} in Fig. \ref{figI}), but ferromagnetic along
$J_{diag}^{<}$ (\emph{red}). Unfortunately, the neutron
measurements suffer from the small vanadium scattering length
causing enlarged error bars for variables which depend on one or
on two V-position parameters. This effect is most visible in Fig.
\ref{figD}~(f) where the difference of the bond lengths of the
diagonal V-V bonds (denoted by the \emph{red} and \emph{green}
lines in Fig. \ref{figI}) is shown. Therefore, an extremely long
measurement has been performed on the D1A diffractometer in order
to increase the statistics, and another synchrotron radiation
experiment at 20~K has been performed at beamline B2 at
Hasylab/DESY. In this synchrotron experiment several image-plate
data collections with a total measuring time of more than one hour
have been summed up. The results of these precise experiments are
plotted together with the other neutron data in Fig.
\ref{figD}~(a-g). The error bars of the synchrotron measurements
are of similar dimensions as the symbol size. Especially, the
value of the difference of the diagonal V-V distances shown in
Fig. \ref{figD}~(f) has much smaller error bars and corroborates
the neutron results. The structural parameters of this synchrotron
measurement and of the neutron measurements at 2~K and at 150~K
are listed in Tab. \ref{tab1}. The difference in the squares of
the diagonal V-V distances is determined with much higher
precision as the V $z$-parameter does not enter its calculation
which only depends on the projection of the distances onto the
$a,b$-planes, see Fig. \ref{figD}~(g). Since the $x$ and $y$
coordinates are fixed, the difference in the squares of the V-V
distances is directly calculated from the lattice parameters.

Indeed, the two diagonal V-V bonds (\emph{red}/\emph{green} in
Fig. \ref{figI}) differ by 0.0073(15) \AA\ as can be seen in Fig.
\ref{figD} (f). The splitting of the diagonal V-V distance is just
a consequence of the monoclinic angle and clearly shows up in
$\sqrt{\{d(V-V)_{diag,1}\}^2-\{d(V-V)_{diag,2}\}^2}$, see Fig.
\ref{figD}~(g). Compared to the orthorhombic structure where the
diagonal V-V bonds  (\emph{red}/\emph{green} in Fig. \ref{figI})
are equivalent, the monoclinic distortion renders  the two
antiferromagnetic exchange interactions inequivalent. We
unambiguously find that the diagonal V-V distance associated with
$J_{diag}^{>}$ is smaller than that associated with $J_{diag}^{<}$
which indicates a stronger orbital overlap accompanied by a larger
exchange interaction $J_{diag}^{>}$  thereby explaining the
observed magnetic structure. Since the magnetic energy gain due to
the lifting of the frustration is linear in the structural
distortion, and since the lattice energy cost scales quadratically
with the distortion, one should always expect such an effect if
the magnetic coupling between two sub-units remains frustrated.
However in most cases, the magnetoelastic coupling would be too
weak to generate an observable effect. The case of VOCl seems to
profit from the fact that the diagonal interaction is modulated
via the V-V distances without variation of the V-O distances. The
structural modulation in VOCl causes thus less lattice energy loss
than for instance a direct bond-distance modulation.

The splitting of the diagonal V-V distances is accompanied by a
small splitting of the V-O-V angles leading to a smaller angle at
the antiferromagnetic bond which seems to contradict the first
estimation that the magnetic interaction should increase with the
bond. However, in this $t_{2g}$ system the direct overlap varied
through the diagonal V-V distance seems to be the dominant effect.

\section{Conclusion}

In conclusion we have studied the nuclear and magnetic structure
of VOCl. Synchrotron diffraction measurements reveal a structural
phase transition $Pmmn$ $\rightarrow$ $P2/n$ which accompanies
the magnetic ordering in VOCl. By neutron diffraction we were
able to refine the magnetic structure within a monoclinic
$2\times2\times2$ unit cell and derived a 3D-antiferromagnetic
structure with an antiferromagnetic moment pointing in
$a$-direction. At 2 K this moment amounts to about 1.3 $\mu_B$
which is a significantly reduced value for a V$^{3+}$-ion
in this 3d$^2$-system.

Magnetoelastic coupling plays an important role in VOCl due to
the fact that the magnetic exchange interaction depends
sensitively on the V-O-V bond angles and on the orbital overlap.
The modulation of the bond angles and of V-V distances involves
less lattice-energy cost than e.g. bond-length modulation, so that
rather strong structural effects occur. First the VO layers
flatten upon cooling which can be well understood in terms of an
enhanced magnetic interaction at low temperatures. The
anisotropic thermal expansion further strengthens the enhanced
magnetic interaction at low temperature. Second, the monoclinic
distortion accompanying the magnetic ordering lifts the
frustration of the coupling between two magnetic subsets as the
diagonal V-V distances shrink in one diagonal direction enhancing
the orbital overlap and, thus, the exchange interaction.

\section{Acknowledgemets.}
This work was supported by the Deutsche Forschungsgemeinschaft
through Sonderforschungsbereich 608.


\begin{thebibliography}{}
\bibitem{tioclshaz} M. Shaz, S. v. Smaalen, L. Palatinus, M. Hoinkis, M. Klemm, S. Horn, and R. Claessen, Phys. Rev. B \textbf{71}, 100405 (2005).
\bibitem{tioclseidel}
A.~Seidel, C.~Marianetti, F.~Chou, G.~Ceder und P.~Lee, Phys. Rev. B {\bf 67},
  20405 (2003).
\bibitem{tioclkataev} V.~Kataev, J.~Baier, A.~M\"{o}ller, L.~Jongen, G.~Meyer und A.~Freimuth, Phys. Rev. B {\bf 68}, 140405 (2003).
\bibitem{tioclrueckamp} R.~R\"{u}ckamp, J.~Baier, M.~Kriener, M.~W. Haverkort, T.~Lorenz, G.~S. Uhrig, L.~Jongen, A.~M\"{o}ller, G.~Meyer und M.~Gr\"{u}ninger, Phys. Rev. Lett. {\bf 95}, 97203 (2006).
\bibitem{tioclcaimi} G. Caimi, L. Degiorgi, N. N. Kovaleva, P. Lemmens, F. C. Chou, Phys. Rev. B \textbf{69}, 125108 (2004).
\bibitem{tioclschoenleber} A. Sch\"{o}nleber, S. v. Smaalen and L. Palatinus, Phys. Rev. B\textbf{ 73}, 214410 (2006).
\bibitem{cugeo1}
M.\ Hase, I.\ Terasaki and K.\ Uchinokura.
\newblock {\em Phys.\ Rev.\ Lett.}, \textbf{70}, 3651 (1993).

\bibitem{tioclvenien} J. P. V\'{e}nien, P. Palvadeau, D. Schleich, J. Rouxel, Mat. Res. Bull. \textbf{14}, 891 (1979).
\bibitem{tioclwiedenmann} A. Wiedenmann, J. P. Venien, P. Palvadeau and J. Rossat-Mignod, J.Phys. C.: Solid State Phys. \textbf{16}, 5339 (1983).
\bibitem{feocl} S. M. Kauzlarich, J. L. Stanton, J. Faber, Jr., and B. A. Averill, J. Am. Chem. Soc. \textbf{108}, 7946 (1986).
\bibitem{feoclB} S. R. Hwang, W.-H. Li, K. C. Lee, J. W. Lynn and C.-G. Wu, Phys. Rev. B \textbf{62}, 14157 (2000).
\bibitem{crocl} A. N. Christensen, T. Johansson and S. Qu\'{e}zel, Acta Chem. Scand. A \textbf{28}, 1171 (1975).
\bibitem{voclhaase} A. Haase and G. Brauer, Acta Cryst. B \textbf{31}, 2521 (1975).
\bibitem{tioclbenckiser} E. Benckiser, R. R\"{u}ckamp, T. M\"{o}ller, T. Taetz, A. M\"{o}ller, A. A. Nugroho, T. T. M. Palstra, G. S. Uhrig and M. Gr\"{u}ninger, New J. Phys. \textbf{10}, 053027 (2008).

\bibitem{cugeo2}
M.\ Braden, G.\ Wilkendorf, J.\ Lorenzana, M.\ A{\"i}n, G.J.\
McIntyre, M.\
  Behruzi, G.\ Heger, G.\ Dhalenne and A.\ Revcolevschi, Phys. Rev. B, \textbf{54}, 1105, (1996).
\bibitem{cugeo3} W.~Geertsma and D.~Khomskii, Phys. Rev. B \textbf{54}, 3011, (1996);
R.~Werner, C.~Gros and M.~Braden, Phys. Rev. B \textbf{59}, 14356, (1999).


\bibitem{tioclschafer} H. Sch\"{a}fer and F. Wartenpfuhl, J. Less. Com. Metals \textbf{3}, 29 (1961).
\bibitem{fullprof} T. Roisnel and J. Rodriguez-Carvajal, FullProf Suite (2008).
\bibitem{beamlineB2A} M. Knapp, C. Baehtz, H. Ehrenberg and H. Fuess, J. Synchrotron Rad. \textbf{11}, 328 (2004).
\bibitem{beamlineB2B} M. Knapp, V. Joco, C. Baehtz, H.H. Brecht, A. Berghaeuser, H. Ehrenberg, H. von Seggern and H. Fuess, Nuclear Instruments and Methods in Physics Research A\textbf{ 521}, 565 (2004).
\end{thebibliography}
\end{document}